\documentclass[a4paper,12pt]{article}
\usepackage{graphicx,rotating,colortbl,xcolor,wrapfig,hyperref,slashed}
\hypersetup{colorlinks,bookmarksopen,bookmarksnumbered,
linkcolor=blus,pdfstartview=FitH,urlcolor=rossos,citecolor=verde}

\def\to{\rightarrow}
\def\bea{\begin{eqnarray}}
\def\eea{\end{eqnarray}}

\def\mpl{M_{\rm Pl}}
\definecolor{Gray}{gray}{0.95}

\newcommand{\kx}{\kappa}
\newcommand{\lx}{\lambda}

\newcommand{\htt}{\tilde{h}}
\newcommand{\Vt}{\tilde{V}}
\newcommand{\Rh}{{R_{\rm h}}}
\newcommand{\Rht}{\tilde{R}_{\rm h}}
\newcommand{\rt}{\tilde{r}}

\newcommand{\Rt}{\tilde{R}}

\newcommand{\Rc}{{\cal R}}
\newcommand{\Kc}{{\cal K}}

\definecolor{rosso}{cmyk}{0,1,1,0.4}
\definecolor{rossos}{cmyk}{0,1,1,0.55}
\definecolor{rossoc}{cmyk}{0,1,1,0.2}
\definecolor{blu}{cmyk}{1,1,0,0.3}
\definecolor{blus}{cmyk}{1,1,0,0.6}
\definecolor{bluc}{cmyk}{1,1,0,0.1}
\definecolor{verde}{cmyk}{0.92,0,0.59,0.25}
\definecolor{verdec}{cmyk}{0.92,0,0.59,0.15}
\definecolor{verdes}{cmyk}{0.92,0,0.59,0.4}

 \def\be   {\begin{equation}}   \def\ee   {\end{equation}}
 \def\ba   {\begin{array}}      \def\ea   {\end{array}}

\font\tenrsfs=rsfs10 at 12pt
\font\sevenrsfs=rsfs7
\font\fiversfs=rsfs5
\newfam\rsfsfam
\textfont\rsfsfam=\tenrsfs
\scriptfont\rsfsfam=\sevenrsfs
\scriptscriptfont\rsfsfam=\fiversfs
\def\mathscr#1{{\fam\rsfsfam\relax#1}}

\def\circa#1{\,\raise.3ex\hbox{$#1$\kern-.75em\lower1ex\hbox{$\sim$}}\,}

\newcommand{\beq}{\begin{equation}}
\newcommand{\eeq}{\end{equation}}

 \def\ex{\epsilon}
 
 \def\kx{\kappa}
 
 \def\lx{\lambda}
 \def\calM{{\cal M}}

\def\rt{{\tilde{r}}}

\def \lta {\mathrel{\vcenter
     {\hbox{$<$}\nointerlineskip\hbox{$\sim$}}}}
\def \gta {\mathrel{\vcenter
     {\hbox{$>$}\nointerlineskip\hbox{$\sim$}}}}

\begin{document}
%CERN-PH-TH-2015/119   \hfill IFUP-TH/2015

\thispagestyle{empty}
\vspace{0.1cm}
\begin{center}

{\Large \bf 
%\color{rossos} 
On the Catalysis of the Electroweak Vacuum Decay  
\\ 
\vskip 0.5cm
by Black Holes  at High Temperature}  \\[2cm]

{\bf D. Canko$^{a}$, I. Gialamas$^{a}$, G. Jelic-Cizmek$^{b}$, 
A. Riotto$^{b}$ and N. Tetradis$^{a}$ } \\[5mm]

\end{center}
\begin{center}
\noindent
{\it 
a) Department of Physics, National and Kapodistrian University of Athens,
\\ Zographou 157 84, Greece
\\
%and 
%\\
b) Department of Theoretical Physics and Center for Astroparticle Physics (CAP),
\\
24 quai E. Ansermet, CH-1211 Geneva 4, Switzerland
%Physics Department, Theory Unit, CERN, CH-1211 Geneva 23, Switzerland
}
\end{center}
\begin{center}
\vspace{2cm}
{\bf\color{blus} Abstract}

\begin{quote}
We study the effect of primordial black holes on the classical
rate of nucleation
of AdS regions within the standard electroweak
vacuum at high temperature. 
We find that the energy barrier for transitions to the 
new vacuum, which determines the exponential suppression 
of the nucleation rate,
can be reduced significantly, or even eliminated completely,
in the black-hole background if the Standard Model Higgs is coupled to gravity through the renormalizable term
$\xi {\cal R} h^2$.
\end{quote}
\end{center}

\newpage
\tableofcontents
%\newpage
\normalsize

%\xxx{New notation, maybe not yet consistently employed.
%Hubble constant = $\Hub$.
%Higgs doublet = $\Hig$.
%Higgs = $h$.}

\section{Introduction}
It is well-known that for the current measured values of the Higgs and top quark masses the Standard Model (SM) effective potential develops an instability. Due to the running of the quartic coupling, the effective potential of the Higgs reaches a maximum and then becomes unbounded from below  at   values of the Higgs field of  about $5\times 10^{10}$ GeV.  Our  electroweak vacuum has a lifetime which is many orders of magnitude larger than the age of the Universe and is therefore stable against decay through quantum tunnelling. 

A pertinent question concerns the fate of  the electroweak  vacuum  during the evolution of the Universe in situations in which gravity plays a pronounced role.
The gravitational background can have a significant effect on the rate of 
vacuum decay, leading to its enhancement or suppression  \cite{deluccia}. 
This issue is crucial for the decay of the electroweak vacuum, because of
the extreme sensitivity of the 
Higgs potential to the Higgs and top masses \cite{instab2,instab,espinosa,rychkov,urbano,urbano2}. 
The fact that the gravitational effects are significant during inflation has
led to an extensive investigation of the stability of the electroweak vacuum
during this era, leading to bounds on the inflationary scale \cite{espinosa,cosmo2,zurek,tetradis,rajantie} in order to avoid a catastrophic
singularity characterized as the AdS crunch \cite{deluccia,tetradis,freivogel}.  

Strong gravitational fields are also sourced by black holes, which raises
the question of electroweak vacuum stability in their vicinity. It has been
argued that black holes can act as impurities enhancing the quantum 
decay rate to a level incompatible with the age of the Universe
\cite{gregory}. It is natural to expect a similar enhancement
for classical transitions to the true Higgs vacuum, induced by the 
high-temperature environment of the early Universe \cite{tetradisbh}. 
This second mechanism
can be explored  through a more intuitive approach, with fewer technical uncertainties than the calculation of the quantum tunnelling rate. 
In any case, the question
whether the existence of primordial black holes \cite{pbh,hawking}
is consistent with the stability of the electroweak vacuum has 
important implications for the compatibility of the Standard Model  
of particle physics and the cosmological model \cite{gorbunov}.

In this paper we explore further the scenario of temperature-induced
vacuum decay in the black-hole background \cite{tetradisbh}. Firstly, we
repeat the analysis using the temperature-corrected effective potential
of the Higgs field, instead of the zero-temperature one employed in 
\cite{tetradisbh}. Secondly, we allow for a nonminimal coupling of the
Higgs field to gravity. This additional coupling can have strong
effects on the action of the bounce in the case of quantum tunnelling,
or the energy of the critical bubble in the case of thermal tunnelling, 
%and consequently on the decay rate, 
because it modifies the effective Higgs mass in regions of large
curvature. Depending on the sign of the coupling, the vacuum decay rate may be
reduced if the mass grows and fluctuations are suppressed, or an
instability may appear if the mass is made to vanish. 
Such behavior is known to occur in the context of inflation
\cite{tetradis} and similar features are expected within the strong gravitational field
of a black hole. 

The paper is organized as follows. In section 2 we derive the equations of motion and the relevant expressions for the energy of the bubble configuration.
%, accounting properly for the boundary terms in the action. 
In section 3 we study the finite-temperature effects on the 
Higgs potential and determine the size of the
barrier to be overcome for vacuum decay in the presence of black holes. In section 4 we 
discuss the stability of the electroweak vacuum
and derive constraints for the allowed range of the nonminimal coupling
of the Higgs field to gravity. In the final section we give our conclusions.

\section{Equations of motion}
Our starting point is the action of the SM Higgs coupled to gravity, where we also allow  the renormalizable term whose strength is parametrized by the coupling
$\xi$
\be
S=\int_\calM d^4x \sqrt{-g} \left(-\frac{1}{2}(\nabla h)^2-V(h) +\frac{1}{16 \pi G}\Rc+
\frac{1}{2} \xi h^2 \Rc \right).
\label{aaction} \ee
The equations of motion are 
%\cite{nucamendi}
\begin{eqnarray}
 \left(\frac{1}{8\pi G}+\xi h^2 \right)\left( \Rc_{\mu\nu}-\frac{1}{2}g_{\mu\nu}\Rc \right)&=&
 \nabla_\mu h \nabla_\nu h -g_{\mu \nu}
\left( \frac{1}{2} (\nabla h)^2 + V(h)   \right) 
\label{aeom1} \\
&+&
2 \xi \left( \nabla_\mu (h \nabla_\nu h)-g_{\mu \nu}\nabla_\lx(h \nabla^\lx h)
\right), 
\nonumber \\
\nabla_\mu \nabla^\mu h +\xi h \Rc  &=& \frac{dV(h)}{dh}.
\label{aeom2}\end{eqnarray}
The use of the equations of motion allows to simplify the
action further. 
In particular, taking the trace of Eq. (\ref{aeom1}) expresses the 
curvature scalar $R$ in terms the scalar field. The derivative of the 
potential can be eliminated through Eq. (\ref{aeom2}). One eventually finds
\be
\Rc \left(\frac{1}{8\pi G}+\xi h^2 \right)=(1+6\xi)(\nabla h)^2 +4 V(h)+
6\xi \, h \nabla_\mu \nabla^\mu h.
\label{Reom} \ee
Substitution in Eq. (\ref{aaction}) gives
\be
S=\int_\calM d^4x \sqrt{-g} \left(V(h) +3\xi \,\nabla_\mu (h \nabla^\mu h)\right).
\label{aactioneom} \ee
The second term in the above expression is reduced to a boundary term upon
integration.

We consider now the SM Higgs field in the presence of black holes that can trigger  the decay of the electroweak vacuum by providing  nucleation sites \cite{gregory}. The decay takes place through the formation of bubbles of the new
vacuum around the black holes. 
The critical bubble is a static configuration, whose profile can be obtained 
by solving the equation
of motion of the Higgs field with appropriate boundary conditions.
The field interpolates between values on either side of
the maximum of the potential. 
The ADM mass for the critical bubble is a measure of the energy barrier that
must be overcome for the field to fluctuate beyond the potential maximum.
In a thermal environment, the ratio of this mass and the temperature 
is expected to determine the exponential suppression of 
the vacuum decay rate.
%The characteristic scale of the fluctuations, which determines the
%dispersion of the Higgs field, is set by the environment. If
%thermal equilibrium is assumed, the scale is given by the temperature.
%However, a high density environment with strong fluctuations also
%affects the Higgs field through the Yukawa couplings to the particles
%that contribute to the density. 

The appropriate metric for our analysis has the form
\begin{eqnarray}
{\rm d}s^2&=&-N(r) \, e^{2\delta(r)} {\rm d}t^2+N^{-1}(r)\, {\rm d}r^2
+r^2 \left({\rm d}\theta^2+\sin^2\theta {\rm d}\phi^2 \right),\nonumber\\
N(r)&=&1-2 \frac{G M(r)}{r}.
\label{metric} 
\end{eqnarray} 
With this ansatz, one has to solve the Einstein
and Higgs-field equations for the static critical bubbles. 
%using the potential (\ref{highTpot}). 
The presence of the horizon at $r=R_{\rm h}$ requires 
appropriate boundary conditions at this point.\footnote{When
the spacetime has a boundary, the consistency of the variational
principle requires the presence of a boundary term \cite{gibbons}
\be
S_b=2 \ex \int_{\partial \calM} d^3 y \sqrt{-g_{\rm ind}} 
\left( \frac{1}{16 \pi G} + \frac{1}{2} \xi h^2 \right) \Kc,
\label{aboundary} \ee
with $(g_{\rm ind})_{\alpha \beta}$ the induced metric on the boundary surface,
$\ex=\pm 1$, depending on
whether the surface is timelike or spacelike, respectively, and
$\Kc$ the trace of its extrinsic curvature. We solve the
equations of motion, starting slightly outside the horizon, assuming 
the presence of such a term with $\ex=-1$.}
The function $N(r)$ vanishes on the horizon, which gives
$2GM(R_{\rm h})=R_{\rm h}.$ 
The quantity $R_{\rm h}/(2G)$ can be identified with the black-hole mass 
$M_{\rm bh}$.
In order to obtain a finite ADM mass for the bubble configuration 
we must avoid a singularity in Eq.
(\ref{eom3}) at $r=R_{\rm h}$. This can be achieved by choosing 
$h'(R_{\rm h})$ appropriately.
We also require that 
$h(r)\to 0$ for $r\to \infty$.
We are interested in
the quantity $\delta M(r)=M(r)-R_{\rm h}/(2G)$.
%, which results from the 
%integration of the energy contributions associated with the bubble. 
The asymptotic value $\delta M_{\rm tot} \equiv \delta M(\infty)$
gives an estimate of the barrier 
associated with the production of the bubble around
a central black hole with horizon radius $R_{\rm h}$.

The characteristic scale of the solutions is set by the largest of 
$h_{\rm max}$, the value of the SM Higgs at the maximum of its potential\footnote{While this definition is not gauge-invariant, 
one could define
$h_{\rm max}$ as $V_{\rm max}^{1/4}$ which, thanks to the Nielsen identity, 
is gauge-invaviant \cite{us}.}  
and the temperature $T$, which are expected to be much smaller than the
Planck scale $\mpl$. This  
leads to a significant
simplification of the setup, resulting from the fact that 
$\delta M(r) \ll R_{\rm h}/(2G)$. 

Since the black-hole mass is the leading
factor determining the gravitational background, we write
$M(r)=R_{\rm h}/(2G)+\delta M(r)$ and keep the leading contributions in $G$.
We find that the equations of motion are reduced to 
\begin{eqnarray}
h''+\left( \frac{2}{r} +\frac{R_{\rm h}}{r(r-R_{\rm h})}
\right) h'&=&\frac{r}{r-R_{\rm h}}\frac{dV(h)}{dh},
\label{eom1} \\
\delta M'&=& 4\pi r^2 \left(\frac{1}{2} \frac{r-R_{\rm h}}{r} h'^2+V(h) \right)\nonumber\\
&+& 4\xi \pi r^2 \left(2 \frac{r-R_{\rm h}}{r} h'^2+2h\frac{dV(h)}{dh}
-\frac{R_{\rm h}}{r^2}h\,h' \right),
\label{eom3}
\end{eqnarray}
with the prime denoting a derivative with respect to $r$. The equation for
$\delta$ is reduced to $\delta'=0$, and we set $\delta=1$.
Avoiding the singularity at $r=R_{\rm h}$ in Eq. (\ref{eom1}) requires that we impose 
\be
h'(R_{\rm h})= R_{\rm h}\, 
 \frac{dV(h(R_{\rm h}))}{dh}.
\label{bound2} \ee
This boundary condition correctly reproduces the 
standard condition $h'(0)=0$ in the absence of a black hole.
Our definition of $\delta M (r)$ imposes  $\delta M (R_{\rm h})=0$.  
The value of $h(R_{\rm h})$ must be tuned so that the condition 
$h(r)\to 0$ for $r\to \infty$ is satisfied.

The form of the above equations demonstrates that the Higgs configuration
is not modified by the nonminimal coupling to gravity, and, therefore, is
independent of $\xi$.
The bubble has a negligible 
effect on the gravitational background, which 
is determined by the black-hole mass to a very
good approximation. Allowing for a nonminimal Higgs coupling to gravity
through a term $\sim  h^2 \Rc$ in the action
does not modify this conclusion, because the curvature $\Rc$ vanishes in
a black-hole background. As a result, all modifications to the background
due to the Higgs are suppressed by $\mpl$.

However, the mass of the bubble configuration
receives a correction proportional to $\xi$. 
It can be expressed as
\be
\delta M_{\rm tot}=
F_1(\Rh)+\xi \, F_2(\Rh),
\label{admot} \ee
with 
\begin{eqnarray}
F_1(\Rh) & = & 4\pi \int_{\Rh}^\infty dr \,
r^2 \left(\frac{1}{2} \frac{r-\Rh}{r} h'^2+V(h) \right)
\label{af1} \\
F_2(\Rh) & = & 4\pi \int_{\Rh}^\infty dr \,
 r^2 \left(2 \frac{r-\Rh}{r} h'^2+2 h\frac{dV(h)}{dh}
-\frac{\Rh}{r^2}h\,h' \right).
\label{af2}
\end{eqnarray}
It is possible to put the above expressions in an alternative form. The equation of motion
(\ref{eom1}) can be written as 
\be
\left(r (r-R_{\rm h})h' \right)'=r^2 \frac{dV(h)}{dh}.
\label{aeom1a}
\ee
Substitution into the expression
(\ref{af1}) gives
\begin{eqnarray}
F_1(R_{\rm h})&=& 4\pi \int_{\Rh}^\infty dr \,\left[
\frac{1}{2} \left( r(r-R_{\rm h})h\, h'  \right)'
-\frac{1}{2}r^2h\frac{dV(h)}{dh}+r^2 V(h)
\right]
\nonumber \\
&=& 4\pi \int_{\Rh}^\infty dr \,r^2\left(
-\frac{1}{2}h\frac{dV(h)}{dh}+V(h) \right).
\label{af1a} \end{eqnarray}
Also, the use of Eq. (\ref{aeom1a}) in (\ref{af2}) gives
\be
F_2(R_{\rm h})= 4\pi \int_{\Rh}^\infty dr \,
\left( 2r(r-R_{\rm h})h\, h' -\frac{1}{2}R_{\rm h}\, h^2 \right)'
=2\pi R_{\rm h}\, h^2(R_{\rm h}).
\label{asurface}
\ee

\begin{figure}[!t]
\centering
$$
\includegraphics[width=0.47\textwidth]{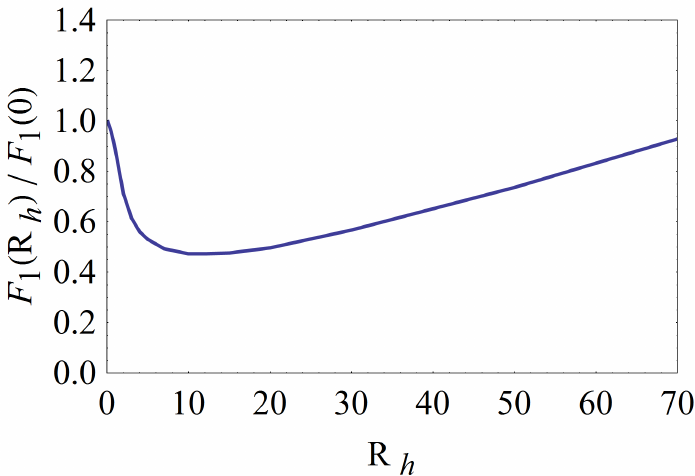}\qquad 
\includegraphics[width=0.47\textwidth]{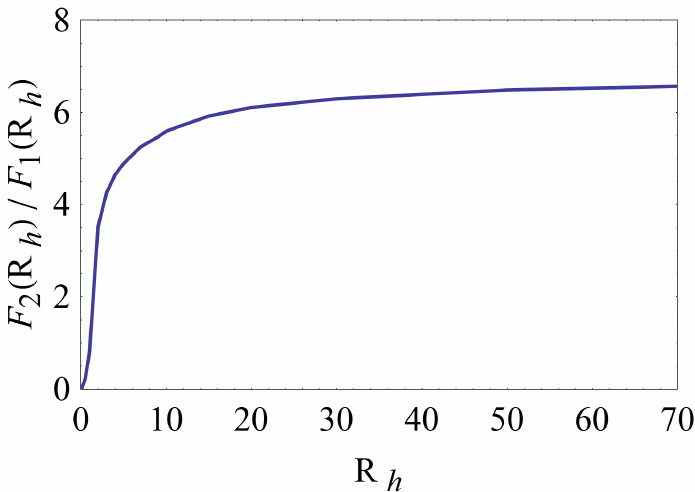}\qquad $$
\caption{\em The quantities $F_1$ and $F_2$, that determine
the bubble mass $\delta M_{\rm tot}$, as a function of the black-hole
radius. All quantities in units of the value of the Higgs field $h_{\rm max}$
at the maximum of the
potential.
}
\label{ffigg}
\end{figure}
In Ref. \cite{tetradisbh} the zero-temperature
 potential was used in order to compute the mass of the bubble configuration
 as a function of the black-hole mass for $\xi=0$. The bubble mass was then used 
 in order to estimate the vacuum decay rate at non-zero temperature. 
In the following section we shall improve on this estimate by considering
the temperature-corrected Higgs potential, which corresponds to the
free-energy density of the system. Before doing so, we present here the 
generalization of the results of Ref. \cite{tetradisbh} for nonzero $\xi$. 
For the Higgs potential we use the approximate form $V \sim \lx(h)\, h^4/4$, 
which is a good approximation for 
values of the Higgs field around its maximum at 
$h_{\rm max}\sim 5\times 10^{10}$ GeV. 
The quartic coupling $\lx$ varies from 0.02 to $-0.02$ 
for Higgs values between $10^{6}$ GeV and $10^{20}$ GeV, respectively. 
 Near the maximum the potential can be approximated as \cite{tetradis}
 \be
 V(h)\simeq -b\ln \left( \frac{h^2}{h^2_{\rm max}\sqrt{e}} \right)\frac{h^4}{4}, \,\,\, b\simeq 0.16/(4 \pi)^2.
 \label{apprvm} \ee 
 In Fig. \ref{ffigg} we plot the functions 
 $F_1$ and $F_2$ that determine
 the bubble mass $\delta M_{\rm tot}$, as a function of the black-hole
 radius. All quantities are measured in units of $h_{\rm max}$.
 We have normalized $F_1(R_{\rm h})$ with respect to its
 value $F_1(0)$ in the absence of a black hole and we depict it in
 the left panel of Fig. \ref{ffigg}.  
 It is apparent that for $\xi=0$ the ADM mass of the
 bubble configuration is reduced by an approximate factor of 2 within a 
 certain range of black-hole masses around 
 $\Rh=10 \, h_{\rm max}$. 
 In the right panel of Fig. \ref{ffigg} we depict the ratio 
 $F_2(R_{\rm h})/F_1(R_{\rm h})$. It vanishes for $R_{\rm h}\to 0$, but
 quickly grows and remains almost constant for $\Rh\gta 20 \, h_{\rm max}$, with 
 a value $\simeq 6.6$. 
  
One may wonder if the problem for $\xi\not= 0$ can be analyzed in the Einstein 
frame as well. There are subtleties in the transition to this frame, related 
to the geometry that we consider. In our analysis we treat the horizon as 
a boundary, assuming the presence of a Gibbons-Hawking term there in order to
make the variational problem well-defined. The reformulation of the problem
in the Einstein frame has to account for both the form of the background 
and the boundary term. We find that our approach is the most transparent.

\section{Finite-temperature effects}
The next step is to consider the immersion of the black hole-Higgs system 
in the thermal environment of the early Universe. This framework raises
certain conceptual issues that need to be clarified. In a thermal environment,
the quantity relevant for transitions between different
states is the free energy, which accounts for the effect of entropy. 
In this respect, it is natural to employ the temperature-dependent 
effective potential, which can be identified with the free-energy density. 
On the other hand, the gravitational field is sourced by
the energy-momentum tensor, which includes the energy density and is 
identified with
the zero-temperature potential. The formal derivation of the appropriate
expressions on a strong gravitational background is difficult. For example,
the effects of temperature are usually taken into account by compactifying
the Euclidean time direction. On a black-hole background, the compactification
scale is set by the Hawking temperature, which does not coincide necessarily 
with the ambient temperature. 

As we saw in the previous section, the backreaction  
of the Higgs field is negligible, so that the background is described
by the Schwarzschild metric to a very good approximation. As a result, the
contributions of the thermal environment to the Einstein equations need not be 
considered. The equations relevant
for our problem, Eqs. (\ref{eom1}), (\ref{eom3}), 
determine the shape and energy of the Higgs 
configuration that the fluctuating system has to go through for the transition
to occur. At nonzero temperature, energy is expected to be replaced by 
free energy.
It is then justifiable intuitively to replace the zero-temperature potential
with the high-temperature one. 
The resulting equations have the correct limits for
vanishing temperature or black-hole mass. We emphasize, however, that
a formal derivation is lacking.

The characteristic scale of the solutions is set by the largest of 
$h_{\rm max}$ and the temperature $T$. 
For $T\gg h_{\rm max}$, the form of the zero-temperature potential near its 
maximum at $h_{\rm max}$ is not relevant for our calculation. The temperature
effects shift the maximum of the potential to a value proportional to $T$. 
The only dimensionful  scale in the problem is the temperature, which 
determines the scale at which the running couplings must
be evaluated. 
The temperature corrections to the Higgs potential are summarized
in Ref. \cite{urbano}. The Higgs field develops a thermal mass 
$m^2_{\rm T}=\kx^2 T^2$, with  
\be
\kx^2 = \frac{1}{12}\left(\frac{3}{4}g'^2+\frac{9}{4}g^2+3y_{\rm t}^2+6 \lx \right)
-\frac{1}{32\pi}\sqrt{\frac{11}{6}}\left(g'^3+3g^3\right)
-\frac{3}{16\pi}\lx \sqrt{g'^2+3g^2+8\lx+4y_t^2}.
\label{kappa2}\ee
The running gauge, $\tau$-Yukawa and quartic running couplings must be evaluated
at a scale set by the temperature.

A semi-analytical treatment of
the effects on thermal tunnelling arising from the 
black-hole background and the nonminimal Higgs-gravity coupling is
possible if we approximate the full potential as
\be 
V(h,T)\simeq \frac{1}{2}\kx^2 T^2 h^2 + \frac{1}{4} \lx h^4.
\label{highTpot} \ee
The quartic $\lx$ is taken to be constant with a value corresponding to
the running coupling at a scale set by the temperature.
In the temperature range between 
$10^{14}$ and $10^{18}$ GeV, we have $\kx \simeq 0.3$, with 
a $10 \%$ decrease for increasing $T$.
Also, $\lx \simeq -0.015$ with a
$15 \%$ decrease for increasing $T$ \cite{urbano}. 
Neglecting the logarithmic running of the quartic coupling is a reasonable
approximation. 
Our discussion can be 
repeated easily for lower reheating temperatures, but a fully numerical 
analysis is necessary if quantitative precision is required.

\begin{figure}
\includegraphics[width=0.7\textwidth]{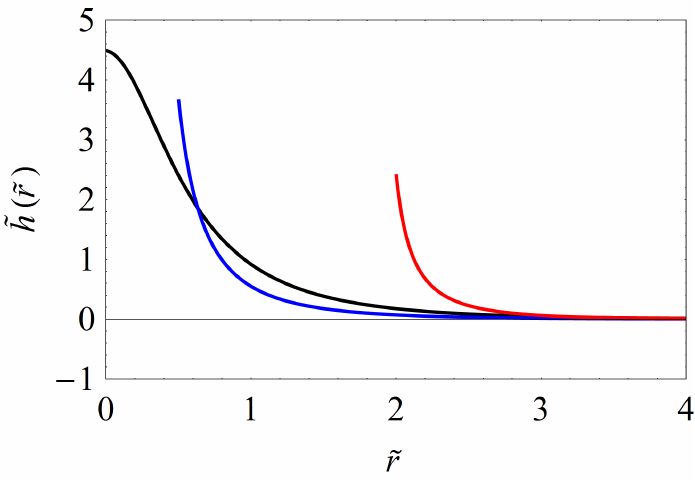}
\caption{The rescaled Higgs field outside a black hole with 
rescaled radius $\Rt_h=0.01,$ 0.5, 2 (lines from left to right).}
\label{fig0}
\end{figure}

%As we have noted, the bubble profile has a characteristic scale set by $T$.
Through the rescalings 
\be
h=\frac{\kappa T}{\sqrt{|\lambda|}} \, \htt, 
~~~~~~~~~~~~~~
r=\frac{1}{\kx T} \, \rt,
\label{rescaling} \ee 
the total free energy of the bubble configuration can be 
expressed as $\kx T/|\lx|$ times a numerical factor
\cite{arnold}. 
Implementing the above,
along with 
\be
\Rh=\frac{1}{\kx T}\, \Rht 
~~~~~~~~~~~~~~
V(h,T)=\frac{\kx^4 T^4}{|\lx|} \Vt(\htt),
\label{rescaling2} \ee 
results in Eq. (\ref{eom1}) being replaced by a similar equation for the
rescaled variables and $\Vt(\htt)=\htt^2/2-\htt^4/4$.
Its solution is presented in fig. \ref{fig0} for three different 
black-hole masses. It is apparent that the characteristic scale of the 
Higgs configuration is comparable to the horizon radius 
for $\Rt_h={\cal O}(1)$. For such black holes, 
the Hawking temperature satisfies
$T_{\rm H}/T=\kx/(4\pi \Rt_{\rm h})={\cal O}(10^{-2})$. There is no
thermal equilibrium between the black hole and the environment, while
the effect of the Hawking radiation on the ambient thermal bath can be
neglected. We shall see below that black holes with $\Rt_h\simeq 0.5$ have
a strong effect on the vacuum decay rate \cite{tetradisbh}. 
This effect is purely gravitational and is localized in the
vicinity of the black hole. As noted in Ref. \cite{gorbunov}, the  
analysis of the quantum decay rate \cite{gregory} results in an expression
that indicates that the compactification of the Euclidean time direction at 
a radius $\sim 1/T_{\rm H}$, linked to the Hawking temperature, 
plays a role at arbitrarily large distances from the
black hole. The effect we are discussing is not of similar nature, and has
a more robust intuitive interpretation.

%\begin{figure}
%\includegraphics[width=0.7\textwidth]{F2.pdf}
%\caption{.}
%\label{fig2}
%\end{figure}
%

The free energy of the bubble configuration can be put in the form
\be
\frac{\delta M_{\rm tot}}{T}=\frac{ \kx}{|\lx|}
\left(F_1(\Rht)+\xi \, F_2(\Rht) \right),
\label{dmot} \ee
with $\kappa/|\lambda|\sim 20$ and 
\begin{eqnarray}
F_1(\Rt_{\rm h})&=&  4\pi \int_{\Rht}^\infty d\rt \,\rt^2\left(
-\frac{1}{2}\htt\frac{d\Vt(\htt)}{d\htt}+\Vt(\htt) \right)=
 4\pi \int_{\Rht}^\infty d\rt \,\rt^2 \, \frac{1}{4}\htt^4.
\label{f1a} 
\nonumber \\
F_2(\Rt_{\rm h})&=&
2\pi \Rt_{\rm h}\, \htt^2(\Rt_{\rm h}).
\label{surface}
\end{eqnarray}
We find $F_1(0)\simeq 18.9$, in agreement with Ref. \cite{arnold}. 
Also $F_2(0)=0$, as can be checked through
a partial integration, use of the equation of motion, and remembering that
$h'(0)=0$ for $\Rh=0$.  Our approximations result in a value for the 
ratio $\delta M_{\rm tot}/T=18.9 \kx/|\lx|$ in the absence of a black hole 
that is independent of the temperature: $\delta M_{\rm tot}/T \simeq 380$.
The complete analysis of the renormalized action of the SM Higgs field 
is consistent with this value, but also indicates a $20 \%$ decrease of the
action for temperatures between $10^{14}$ and $10^{18}$ GeV \cite{urbano,urbano2}. At lower temperatures, the logarithmic running of 
the quartic term in the potential becomes important and numerical 
determination of the bubble profile is needed for every value of $T$. However,
we expect that the effect of the black hole on the ratio
 $\delta M_{\rm tot}/T$ can be described again by 
a multiplicative factor $(F_1(\Rht)+\xi \, F_2(\Rht))/F_1(0)$, 
as in eq. (\ref{dmot}). This expectation is supported by the analysis of
Ref. \cite{tetradisbh}, in which the zero-temperature potential was used.

\begin{figure}[!t]
\centering
$$
\includegraphics[width=0.47\textwidth]{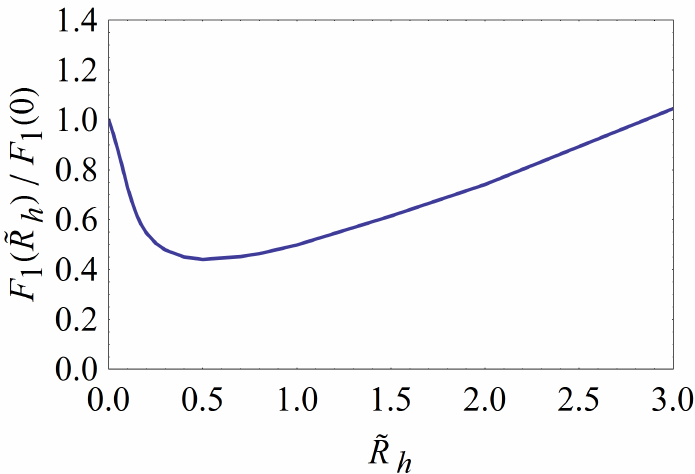}\qquad 
\includegraphics[width=0.47\textwidth]{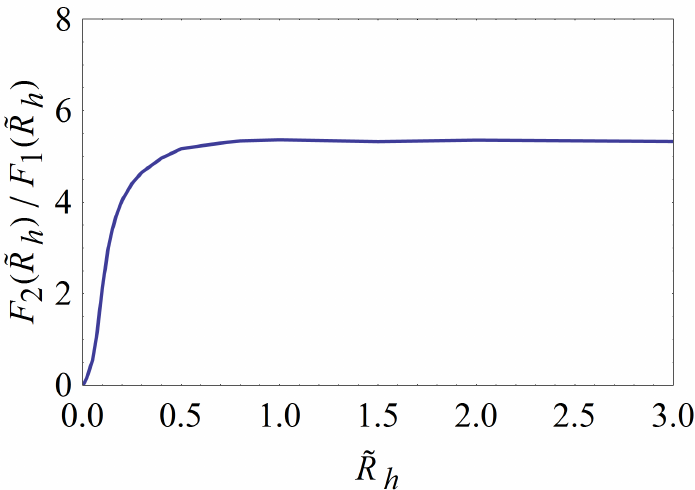}\qquad $$
\caption{\em The dimensionless quantities $F_1$ and $F_2$, that determine
the ratio $\delta M_{\rm tot}/T$, as a function of the rescaled black-hole
radius.
}
\label{figg}
\end{figure}

In Fig. \ref{figg} we depict the functions $F_1(\Rt_{\rm h})$ and 
$F_2(\Rt_{\rm h})$. We have normalized $F_1(\Rt_{\rm h})$ with respect to its
value $F_1(0)$ in the absence of a black hole. 
The form of $F_1(\Rt_{\rm h})$ is similar
to that observed in Ref. \cite{tetradisbh}: the free energy of the
bubble configuration for $\xi=0$ 
is reduced by an approximate factor of 2 within a 
certain range of black-hole masses around 
$\Rh=0.5/(\kappa T) \simeq 1.7/T$. As a result, the nucleation rate for
$\xi=0$ can be reduced significantly in the black-hole background. 
In the right panel of Fig. \ref{figg} we depict the ratio 
$F_2(\Rt_{\rm h})/F_1(\Rt_{\rm h})$. It vanishes for $\Rt_{\rm h}\to 0$, but
quickly grows and remains almost constant for $\Rt_{\rm h}\gta 0.5$, with 
a value $\simeq 5.3$.

A very interesting feature is the effect of the nonminimal coupling
on the bubble free energy. The dependence on $\xi$ is linear, as can be seen 
from Eq. (\ref{dmot}). Large positive values of $\xi$ result in the growth
of $\delta M_{\rm tot}/T$ and the suppression of the nucleation rate.
As we mentioned in the introduction, the nonminimal coupling can have a strong
effect because it modifies the effective Higgs mass in regions of large
curvature. Depending on the sign of the coupling, the rate may be
reduced if the mass grows and fluctuations are suppressed, or an
instability may appear if the mass is made to vanish. 
Such behavior has been discussed in the context of inflation
\cite{tetradis}.

From Eq. (\ref{dmot}) it is apparent that, for every black-hole mass 
at a given temperature, there is a critical negative value 
$\xi_{\rm cr}=-F_1(\Rt_{\rm h})/F_2(\Rt_{\rm h})$, 
for which the exponential suppression of the nucleation rate is eliminated.
Clearly, the saddle-point approximation breaks down before this 
point is reached. On the other hand, the probability of vacuum decay in
the early Universe depends also on the total number of primordial black
holes that can be generated. In the following section we discuss this 
point in detail.

%of $\xi$, given 
%by $\xi_{\rm cr}=-F_1(\Rt_{\rm h})/F_2(\Rt_{\rm h})$, for which 
% $\delta M_{\rm tot}$ vanishes and the exponential suppression of the
% rate disappears. 
% For $\Rt_{\rm h}\to 0 $, the critical value $\xi_{\rm cr}$ diverges,
% but for  $\Rt_{\rm h}={\cal O}(1)$ we have $\xi_{\rm cr}={\cal O}(10^{-1})$. 
%

\section{Vacuum (in)stability in the presence of black holes}

The nucleation probability per unit time in the background of 
a black hole during the radiation-dominated era of the early Universe
is $dP/dt \sim T \exp(-\delta M_{\rm tot}/T)$.
There is no volume factor because of the absence of translational 
invariance. 
The characteristic time interval that can be associated with the 
temperature scale $T$ is the Hubble time $1/H\sim \mpl/T^2$. 
The nucleation can be efficient over longer times, but we are interested
in a lower bound for the probability. 
Neglecting the evaporation of the
black hole, we have a bubble-nucleation probability 
$P\sim \mpl/T \exp(-\delta M_{\rm tot}/T)$. 
The most uncertain factor in our discussion is the number of primordial
black holes within our past light cone.
The number $N$ of causally independent regions at a time during the
radiation-dominated era with temperature $T$, which are currently 
within our horizon, is $N \sim 10^{34} (T/{\rm GeV})^3$ \cite{tetradisbh}.
One needs to estimate the probability $p$ for a black hole to exist within
one of these regions. This is a strongly model-dependent quantity and
little progress can be achieved without assuming a specific mechanism
for black-hole creation. In any case, 
putting everything together, we find that the
logarithm of the 
probability of electroweak vacuum decay in the presence of black holes 
of typical radius $\Rh$ can be written as 
\be
\ln (N p P)=205+ 2 \ln \left(\frac{T}{\mpl} \right)+ \ln p - \left(\frac{\delta M_{\rm tot}}{T} \right)_0 
A(\Rh T) \left[ 1+\xi\, B(\Rh T) \right],
\label{bubnprob} \ee
where 
$\mpl=(8\pi G)^{-1/2}\simeq 2.43 \times 10^{18}$, 
$(\delta M_{\rm tot}/T)_0$ is computed in the absence of black holes, and
$A(\Rh T)\equiv F_1(\Rt_{\rm h})/F_1(0)$, 
$B(\Rh T)\equiv F_2(\Rt_{\rm h})/F_1(\Rt_{\rm h})$
are depicted in Fig. \ref{figg}.

In Ref. \cite{tetradisbh} it was pointed out that $A(\Rh T)$ can
lead to the reduction of the effect of  $(\delta M_{\rm tot}/T)_0$
by a factor of roughly 2. Our present analysis demonstrates that the
nonminimal coupling $\xi$ to gravity can have a more dramatic effect:
For $\xi=-1/B(\Rh T)$, the
exponential suppression of the nucleation rate can be eliminated completely. 
Clearly, the saddle-point expansion around the bubble configuration breaks
down before this point is reached. On the other hand, the presence of a 
large prefactor, coming from the huge number of causally independent
regions in the early Universe, indicates that the electroweak vacuum
is likely to become totally unstable. 

As we have mentioned above, the largest uncertainty in the 
calculation is connected with the probability $p$ to find a black hole within
each causally independent region at the time when the ambient temperature is
$T$. A primordial 
black hole can form when the density fluctuations are sufficiently large
for an overdense region to collapse \cite{pbh}. 
It is usually assumed that its typical mass is of the order of its maximal possible mass. The latter is given by the total mass within the 
particle horizon $\sim \mpl^2/H$,
while the maximal radius is $\sim 1/H$. It is not clear, however,
if these assumptions are consistent with the typical size of density 
fluctuations in the early Universe, or the constraints imposed by the
observations of the microwave background, for example.

As a concrete application, we consider the scenario of 
Ref. \cite{gorbunov}, which is consistent with the
current bounds on the size of primordial density fluctuations. It is assumed
that the inflationary era is followed by a period during which the inflaton
oscillates and decays into particles. 
The effective equation of state is similar to that
of a matter-dominated Universe. It is possible, therefore, for perturbations
that were generated during inflation to reenter the horizon, grow gravitationally
and collapse into black holes. When the thermalization of the decay products takes place, 
the equation of state changes and the growth of perturbations is suppressed. 
In such a scenario, the black holes most relevant for our discussion are those produced just before thermalization, because they are the most massive ones.
This must be contrasted with the scenario of the quantum decay of the 
electroweak vacuum in which the most relevant black holes are the light ones, because of their large Hawking temperature \cite{gregory,gorbunov}. 

The modes that result in black-hole creation start from a primordial 
value $\delta \rho/\rho = \delta_i\sim 10^{-4}$ at horizon crossing and
grow until  $\delta \rho/\rho \sim 1 $, at which point they decouple from 
the Hubble flow, turn around and collapse. 
The probability for them to form black holes has been estimated as
$P_{\rm BH} \sim 2\times 10^{-2} \left( {R_{\rm h}}/{R}  \right)^{13/2}$
\cite{khlopov},
with $\Rh$ the Schwarzschild and $R$ the turnaround radius.
The black-hole mass is equal to that contained within the turnaround radius:
$M=4\pi \rho R^3/3$. The Friedmann equation for a matter-dominated era then
gives: $\Rh/R \simeq H^2 R^2 \simeq \delta_i$. We can also deduce that
$\Rh\simeq H^2 R^3 \simeq \delta_i^{3/2}/H$. As one expects $\sim (HR)^{-3}$
regions of size $R$ within a volume $\sim 1/H^3$, the probability to find a black hole within the horizon is \cite{gorbunov}
\be
p \sim 2\times 10^{-22} \left( \frac{\delta_i}{10^{-4} }  \right)^{5}.
\label{probbhhor} \ee
The bubble-nucleation rate is enhanced if, at the onset of the
radiation-dominated era, there are black holes for which 
$\Rht=\kx \Rh T \gta 0.5$.  At that time 
$\rho=(g_* \pi^2/30) T^4$, where we take $g_*=106.75$, assuming only 
the particle content of the Standard Model. We obtain
\be
\Rht=\kx \Rh T \simeq \kx \left(\frac{90}{g_* \pi^2} \right)^{1/2} \,\delta_i^{3/2}\,
 \frac{\mpl}{T}  \simeq 0.1 \, \delta_i^{3/2}\, \frac{\mpl}{T}.
\label{reht} \ee
If the reheating temperature is larger than $\sim 10^{-7} \mpl$,
$\Rht$ is too small at the onset of the radiation-dominated era for the black holes
to have a siginificant effect on the rate. At later times, black-hole 
creation is suppressed, the temperature
drops, while the black-hole mass is reduced through evaporation. 
As a result $\Rht$ becomes even smaller. 

It seems, therefore, that a high
reheating temperature eliminates the danger posed by the presence of black holes.
On the other hand, for $T \gta 10^{-7} \mpl$, the typical mass of a black hole
in this scenario is $M \lta 10^{27}$ GeV $\simeq 10^3$ gr. Such small black holes
evaporate very quickly and are of little phenomenological interest. 
In this respect, the possibility of a reheating temperature smaller than
$10^{-7} \mpl$ seems more exciting. 
As an example, let us consider the possibility of a reheating temperature 
$T\simeq 5\times 10^{11}$ GeV, for which Eq. (\ref{reht}) with $\delta_i \simeq 10^{-4}$ gives 
$\Rht \simeq 0.5$, so that $A(\Rh T)\simeq 0.5$ and $B(\Rh T)\simeq 5.3$ 
in Eq. (\ref{bubnprob}). Using Eq. (\ref{probbhhor}), we obtain
$\ln (N p P)\simeq 124-0.5 (1+5.3 \, \xi)(\delta M_{\rm tot}/T)_0$, 
with $(\delta M_{\rm tot}/T)_0$ computed in the absence of black holes.
The probability becomes of order one for 
$\xi \lta -0.19+47/(\delta M_{\rm tot}/T)_0$. The stability of the
electroweak vacuum in the presence of primordial black holes imposes 
a strong constraint on the nonminimal coupling of the Higgs field to
gravity, by forbidding this range.

\section{Conclusions}
Given the absence of new physics in the LHC data, the  stability of the 
SM electroweak vacuum has become a more pressing  issue.
Special conditions in the 
early Universe may not  be  left out when 
studying the rate of vacuum decay. As a particular example, 
black holes may trigger the decay, as first suggested in Ref. \cite{gregory}. 
In this paper  we have analyzed 
how the presence of primordial black holes may induce the SM electroweak vacuum 
decay in the early Universe at finite temperature. In particular, our 
findings indicate that a nonminimal, but renormalizable coupling between the SM 
Higgs field and gravity  may alter considerably the decay rate. 

Ultimately, 
the final fate of the electroweak vacuum is a model-dependent issue, which 
suffers from our ignorance of the precise early Universe dynamics and the 
exact black hole mass function as a function of time. Nevertheless, our results 
indicate that even moderate values of the coupling $\xi$ between the Higgs and 
gravity can render more stable or unstable the electroweak vacuum, depending on 
the sign of the coupling. While this calls for a refinement of our analysis, 
e.g. by determining more precisely 
the probability $p$ for a black hole to exist in one of 
the many causally independent 
regions which are currently within our visible Universe, 
it demonstrates once more the
importance of gravity for the issue of vacuum decay.

\subsubsection*{Acknowledgments}
N.T. would like to thank A. Salvio, A. Strumia and A. Urbano for 
useful discussions.
 A.R. is supported by the Swiss National Science Foundation (SNSF), project {\sl Investigating the
Nature of Dark Matter}, project number: 200020-159223.

    \end{document}